# Investigation of liquid lines as terahertz emitters under ultrashort optical excitation

Qi Jin,[1] Yiwen E,[1] Shenghan Gao,[1] and Xi-Cheng Zhang[1,2,a]

[1]*The Institute of Optics, University of Rochester, Rochester, NY 14627, USA*

[2]*Beijing Advanced Innovation Center for Imaging Technology, Capital Normal University, Beijing 100037, China*

[a] Corresponding author: xi-cheng.zhang@rochester.edu

Recently, there has been growing interest in terahertz (THz) wave generation from liquids under optical excitation. Here, we propose and demonstrate the use of liquid lines in place of liquid films as THz emitters to boost THz signals. The emitter's geometry eliminates the total internal reflection at the flat liquid/air interface. Optimization of optical pulse duration and liquid line's diameter are also experimentally investigated. In addition, we observe that the polarity of liquid has a significant influence on the THz wave generation. α-pinene, a nonpolar liquid, offers much stronger THz radiation than water does. Besides paving the way to develop intense liquid THz sources, our work indicates that THz waves could be a tool for the further study of laser-liquid interaction.

Due to intriguing applications in nonlinear optics[1,2] and electron acceleration,[3,4] development of laser-based intense terahertz (THz) sources becomes a highly influential topic in the community. It has been demonstrated that gas plasmas[5-7] and nonlinear crystals[8-11] are capable of radiating intense THz waves under the excitation of laser pulses.

Recently, THz wave generation from liquids has also been demonstrated.[12,13] A ponderomotive force-induced dipole model has been applied to explain the observations.[14] Ionized electrons experience the ponderomotive force in liquid and accelerates backward. A dipole created along the direction of laser propagation radiates broadband electromagnetic waves including THz frequencies. Additionally, a two-color excitation scheme has been applied to greatly enhance the THz radiation in liquid.[15] Compared with gas, liquid has a lower ionization threshold but higher molecular density,[16-19] which indicates that liquid is capable of providing more electrons and ions in the same unit volume. Unlike nonlinear crystals and organic crystals that suffered from permanent damage under the excitation of intense laser pulses, [8,10,20] liquid can quickly replenish itself due to its fluidity. Moreover, THz waves emitted from liquid can be scaled up by increasing the pump laser pulse's energy while no saturation effect has been experimentally observed[12,14,15]. Hence, even though liquids are only capable of generating kV/cm level THz field by so far, they have the potential to be developed as intense THz sources to meet desired applications.

As the most ubiquitous liquid on the earth, water attracts the attention of being a THz source. To compromise strong absorption of liquid water in the THz frequency region,[21] a wire-guided free-flowing thin water film has been developed to emit THz waves under the excitation of laser pulses.[12] It is observed that the film will break if high-energy laser pulses are focused into it. To overcome this, a nozzle jet was used to produce a water film to bear intense focused laser pulses in ref. [14]. However, a common problem occurred in the film geometry is that the majority of generated THz waves cannot be coupled out of the water film. Specifically, THz signals are strongly absorbed by the millimeter-width water along the side-way directions. Besides the absorption of water, the total internal reflection at the flat water/air interface will significantly reduce THz signals.[14] Furthermore, due to the limitation by the fixed dimension of nozzle jets, water films with different thicknesses are not easily attainable, which impedes the systematical investigation of the optimization conditions and mechanisms of intense THz wave generation from liquid water.

In this letter, liquid lines are introduced to ameliorate the drawbacks mentioned above for THz wave generation from liquid films. Fig. **1** captures a 260 μm diameter water line, which is produced by a 260 μm inner diameter syringe needle. The flow velocity is controlled to be 7 m/s. In this case, each laser pulse will interact with a fresh water spot. We define that the water line is flowing along the y-direction. A femtosecond amplified laser with a central wavelength of 800 nm and a repetition rate of 1 kHz is used for the excitation. The horizontally polarized laser beam with pulse energy of 0.8 mJ propagating along the z-direction is focused into the water line by a 2-inch effective focal length lens (F/4) to generate THz waves. A high-resistivity silicon wafer acts as the filter to block the residual laser beam while allowing the THz beam passing through. THz electric fields are measured by a 2-mm thick, ⟨110⟩-cut ZnTe crystal through electro-optical sampling (EOS),[22] which is placed in the direction of laser propagation.

Figure **2a** shows the peak values of THz fields when the 260 μm diameter water line is scanned along the x-direction across the laser focal point. The scanning direction, water flowing direction, and laser propagation direction are perpendicular to each other. Weak THz signal is detected at x = 0 μm, which is represented as the black dot in the middle. This coincides with the observation in the case of the water film with a normal incident laser beam,[14] where the dipole is along the direction of laser propagation. Non-zero THz signal is detected from a tilted dipole when the water

line is shifted away from the zero position in the x-direction. The THz field is maximized at x = ± 90 μm, which could be a combined result of laser-water interaction and THz absorption of water. The values of THz peak fields change the sign from negative (red dots) to positive (blue dots) while the x position varies from negative to positive. As one can see, if two points are symmetric around 0, their absolute values of THz peak fields are almost identical. The temporal waveform of THz signals at x = ± 90 μm is shown in Fig. **2b**. Clearly, the THz waveform flips over when their x positions are symmetric with respect to the zero position. The observations in Fig. **2** are well predicted and explained by the dipole model approximation proposed in ref. [14]. Opposite x positions lead to opposite projected directions of generated dipoles in water, resulting in upside-down THz waveforms. For x = 0 μm, the dipole in water generated by focused laser pulses along the direction of laser propagation contributes little THz radiation in the z-direction. Thus, a tilted radiating dipole realized by a shift of the water line in the x-direction is essential to obtain strong THz signals in the direction of laser propagation. We need to mention that unlike the case of water film, THz signals generated by a water line can be coupled out and detected in the x-direction. This will increase the overall coupling of THz radiation.

It is noticed that the thicknesses of water film used in ref. [12] and ref. [14] are different. The laser pulse duration is individually optimized in these two cases. This phenomenon reveals the speculation that optimized laser pulse duration depends on the effective path length (EPL) in water. The two main ionization processes in liquids are multiphoton ionization and cascade ionization.[18,19,23] Multiphoton ionization prefers shorter optical pulse duration while cascade ionization prefers longer optical pulse duration. With the increasing volume of water, cascade ionization become the dominated ionization process by its capability of offering an exponential increase in the number of electrons. Therefore, the optimization will be in favor of a long pulse duration with a thicker water film. The data from a series of different water film thickness are indispensable to draw a conclusion. Benefited from the availability of a series of syringe needles with various sizes, water lines with diameters varied from tens of microns to a few millimeters are realized. In our experiment, seven individual syringe needles with different inner diameters are used to produce water lines. Fig. **3a** plots the optimized laser pulse duration as a function of the water line diameter. The optimized pulse duration gradually changes from 273 fs to 546 fs when the diameter of the water line increases from 90 μm to 510 μm. As expected, a longer optical pulse duration is required to optimize the THz signal when the diameter of the water line increases.

As shown in Fig. **3b**, the maximum THz energy is obtained with a 210 μm diameter water line, which results from the trade-off between laser-water interaction length/volume and THz absorption of water. Specifically, a very thin water line does not have sufficient length/volume for the interaction between laser pulses and water molecules while a very thick water line suffers from serious THz absorption of water. Either case does not offer strong THz radiation. It needs to be mentioned that the THz energy slightly decreases rather than significantly drops when the diameter of the water line increases from 330 μm to 510 μm. This observation seems to go against the exponential decay caused by the absorption of water. In fact, during the optimization process for the THz signal, the water line was moved toward the incoming laser beam along the y-direction to compensate for the absorption of water. Consequently, the laser focal point was close to the second interface (water/air interface) instead of the center of the water line. The result also reveals that further increasement in the diameter of the water line does not help anymore once the interaction length is long enough.

The majority of observations in THz wave generation from water[12,14,15] assemble those from the air because laser-induced breakdown and ionization play crucial roles in both generation processes. For the one-color optical excitation scheme, the model based on a ponderomotive force-induced dipole was commonly used in the case of water and air[14,24,25]. For the two-color optical excitation scheme, the transient current model was successfully applied to both cases[15,26-28]. Remarkably, results in Fig. **3** are only found in water rather than in air. Besides interfaces induced optical refraction/THz reflection and bulk water induced absorption, laser-water interaction is also a key factor leading the unique observation in water. It implied that laser-induced breakdown and ionization in the water are different from that in the air. Regarding the THz wave generation from water, complicated nonlinear processes[29-31] (e.g. self-focusing, intensity clamping, self-steepening, and spectral broadening) are also involved in the laser-water interaction. To develop a complete model that addresses the whole generation process, further experimental and theoretical investigations are needed.

Figure **4** plots THz waveforms generated by a 210 μm diameter water line and a 120 μm thick water film. The angle of incidence on the water film is 60°, which leads to the EPL = 158 μm in water. Even though the shift of the water line in x-direction makes the EPL to be 151 μm, Fig. **4** still shows a reasonable comparison. 2.8-times enhanced THz electric field (peak-valley) is obtained by using the water line to substitute the water film. Comparing with the flat water/air interface of a water film, the curved interface of a water line helps to reduce the total internal reflection of THz waves, and therefore, more THz radiation can be coupled out.

It is noteworthy that the current experimental setup only needs 20 mL liquids running in a closed-loop circulation system. The low consumption enables investigations of THz wave generation from the diverse liquid family. One important property of liquid is polarity, which is related to the absorption of THz waves: polar liquids strongly absorb THz radiation while nonpolar liquids do not.[32] Polarity index is a quantitative parameter that defines how polar a liquid is.[33] Stronger THz signals are expected from liquids with small polarity indices. Fig. **5a** shows the peak-valley values of THz electric fields obtained from four liquid lines that use α-pinene, p-xylene, ethanol, and water. Their polarity indices are 0 (nonpolar), 2.4, 5.2, and 9 respectively.[33]

From the result of these four liquids, α-pinene offers the strongest THz field, which is about 1.75-times stronger than that from water in the identical experimental condition (liquid line diameter, laser pulse energy, laser pulse duration, F/#, etc.). The electric field strength of THz waves generated by the 210 μm diameter α-pinene line is estimated to be 0.7 kV/cm through EOS.[8] Fig. **5b** plots the comparison in the frequency domain. Among the four liquids, water generates the THz signal with much less high-frequency component. This may due to the stronger absorption of water in the higher THz frequency region.[34] Thus, although ethanol exhibits a peak-valley value of the THz field similar to water, it generates more high-frequency THz radiation. In Fig. **5b**, the widest bandwidth is achieved by p-xylene while the highest amplitude is attained by α-pinene. Even though there are some other factors may also influence the THz radiation from different liquids, such as refractive index, surface tension, and viscosity, the result in Fig. **5** shows that polarity of liquid is significant for the generation of THz waves.

In summary, liquid lines produced by syringe needles achieve stronger THz radiation by mitigating the total internal reflection at the liquid/air interface than liquid films do. It is predictable that the overall coupling is further increased because THz signals in the x-direction can also be coupled out in a liquid line. Owing to the availability of a series of different sizes of liquid lines, we find that the optimized laser pulse duration highly depends upon the diameter of the liquid line, which also impacts on the THz radiation energy. The maximum THz signal is obtained with a 210 μm diameter liquid line. The unique phenomenon in liquid possibly makes THz wave a great tool for the further study of laser-liquid interaction. Additionally, we show that polarity index is a consequential parameter in THz liquid photonics. α-pinene, as a nonpolar liquid, offers 1.75-times stronger THz electric field than water. Our observations contribute enthralling insight into the development of intense liquid THz sources.

Acknowledgement

The research at the University of Rochester was sponsored by the Army Research Office under Grant No. W911NF-17-1-0428 and the Air Force Office of Scientific Research under Grant No. FA9550-18-1-0357.

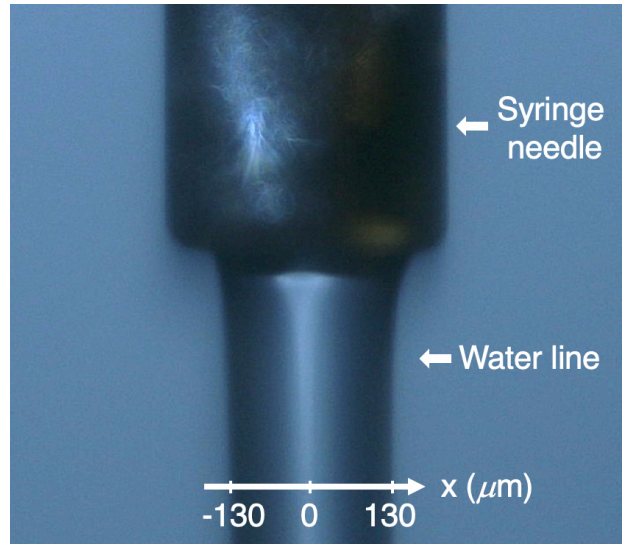

Figure 1. A photo of the water line produced by a syringe needle. The diameter of water line is 260 μm. The flowing velocity is about 7 m/s.

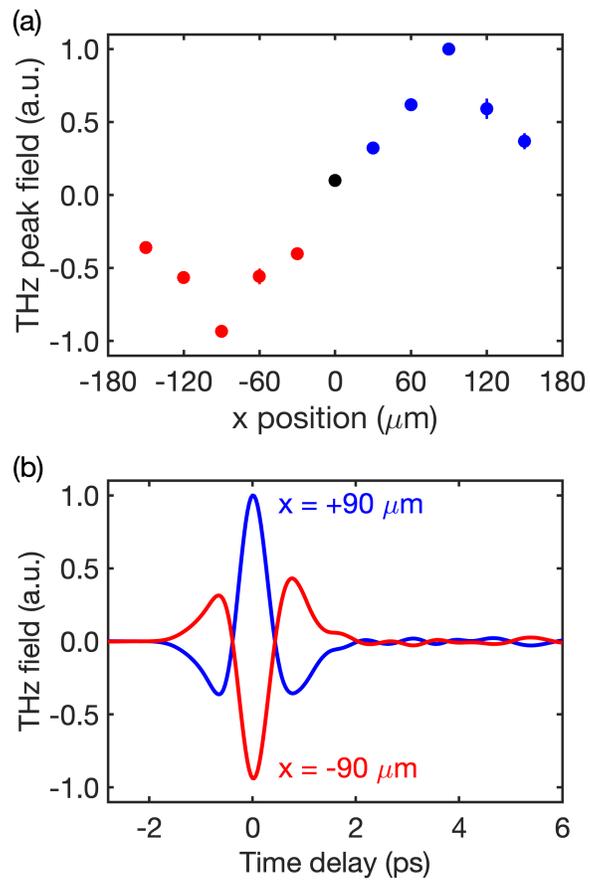

Figure 2. **a** THz peak fields when a 260 μm diameter water line is crossed the laser focal point along the x-direction. **b** THz waveforms of x = ± 90 μm in **a**.

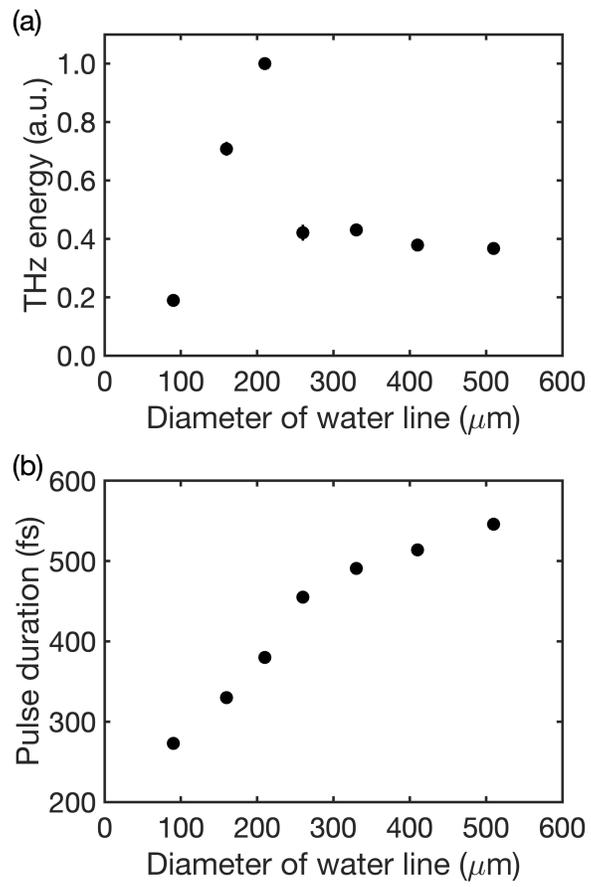

Figure 3. **a** Optimized optical pulse duration as a function of the diameter of water line. **b** THz energy as a function of the diameter of water line.

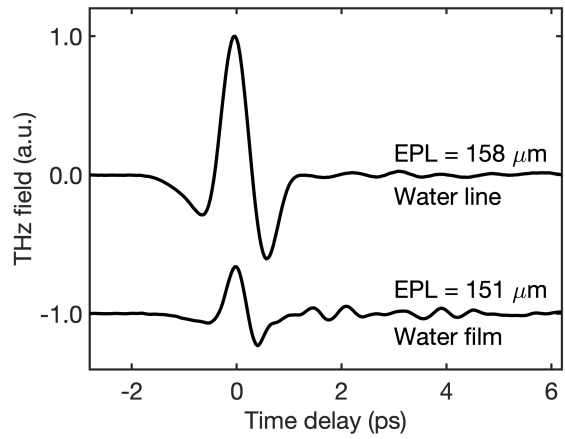

Figure 4. Comparison of THz fields generated from a 210 μm water line and a 120 μm water film. EPL, effective path length. A shift of the 210 μm water line in the x-direction makes the EPL to be 158 μm. Oblique incidence on the 120 μm water film makes the EPL to be 151 μm.

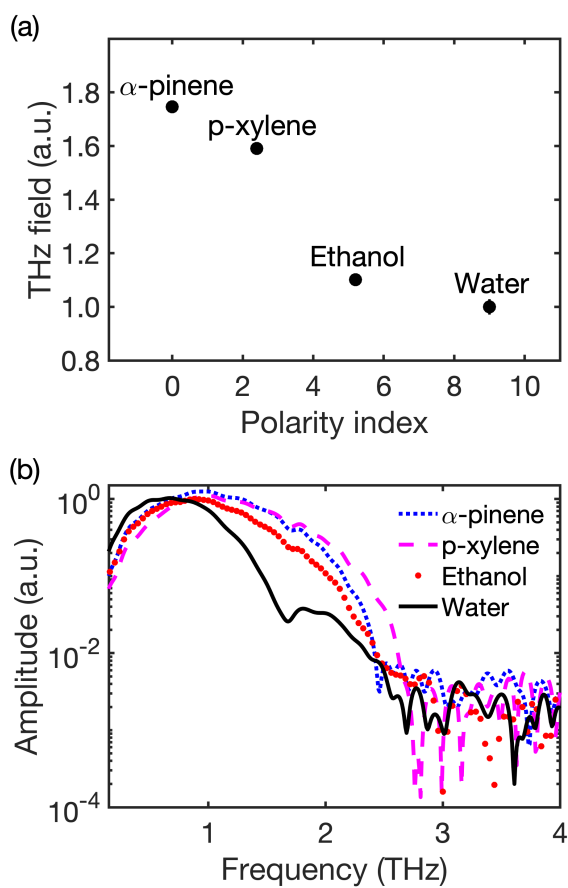

Figure 5. THz fields generated from four liquid lines. The liquids are α-pinene, p-xylene, ethanol, and water. Their polarity indices are 0, 2.4, 5.2, and 9 respectively. The liquid line diameters are 210 μm. Laser pulse energy is 0.8 mJ with optimized pulse duration around 380 fs.